\documentstyle[twocolumn,aps,epsfig]{revtex}
\begin{document}
\draft
\title{Entropy Production and Cosmological Inflation}
\author{T. Sakai, A. Widom and Y.N. Srivastava}
\address{Physics Department, Northeastern University, Boston MA 02115}
\maketitle

\begin{abstract}
The notion of inflation (past or present) in standard cosmological 
models is shown to be a consequence of a sufficiently high second 
law entropy production from the internal heating of the universal  
expansion. The longitudinal viscous internal heating of matter requires 
neither ``inflaton'' fields nor ``quintessence'' fields which in 
theory may induce a cosmological term into the Einstein equations. 
The purely thermodynamic principles required to understand inflation 
within the context of the standard  general relativity equations 
will be discussed in detail.
\end{abstract}  

\narrowtext

\section{Introduction} 

Under the hypotheses of the standard cosmological model one reduces 
space and time scales of the history of the universe to a single central 
function of time \begin{math} a(t) \end{math}\cite{1,2,3,4}. In detail, 
the standard cosmological space-time metric reads 
\begin{equation}
c^2 d\tau ^2=c^2 dt^2-
a^2\left(d\chi^2 +
\sigma_\kappa (\chi )(d\theta ^2+\sin^2\theta d\phi^2)\right),
\end{equation}
where (for \begin{math} \kappa=-1,0,1 \end{math})
\begin{equation}
\sigma_1(\chi )=\sin \chi ,\ \sigma_0(\chi )=1, 
\ {\rm and}\ \sigma_{-1}(\chi )=\sinh \chi.
\end{equation}
The equation of motion for \begin{math} a(t) \end{math} reads 
\begin{equation}
\ddot{a}(t)+\Omega^2(t)a(t)=0,
\end{equation}
where 
\begin{equation}
\Omega^2=\left({4\pi G\over 3 c^2}\right)
(\varepsilon +3P).
\end{equation}

In Eqs(3) and (4) one confidently writes 
\begin{math}\Omega^2 \end{math}, {\em very sure} in the knowledge that the 
energy per unit volume \begin{math}\varepsilon \end{math} and the 
pressure \begin{math}P \end{math} are {\em both positive}. Actually, all one  
knows from the relativistic stability of matter\cite{1,2,3,4} is that 
\begin{equation}
\varepsilon \ge 3P.
\end{equation}  
When it turns out that energy density 
\begin{math}\varepsilon \end{math} and/or the 
pressure \begin{math}P \end{math} are {\em not} positive, one may either 
invent new funny sounding quantum fields, e.g. the ``inflaton field'' or 
the ``quintessence field''\cite{5,6}, or take comfort in what Einstein himself 
called his big research blunder, i.e. the cosmological term\cite{7,8,9,10} 
in the gravitational field equations. Armed with these possible 
new quantum fields, one finds that \begin{math}\Omega^2 \end{math} 
can indeed be negative, at least for {\em some} time periods in the history 
of the universe. Thus, {\em inflation} could exist
\cite{11,12,13,14,15,16,17,18,19,20,21} in {\em very ancient 
epochs}. Recently it has even been suggested, on the basis of very distant 
super nova explosions, that {\em presently} we live in an ancient epoch, 
although the inflation turns out to be rather small. The small 
present inflation is thought to require merely a quintessence field 
(whatever that may be). 

Our purpose is replace the invention of these new quantum fields with 
a more close scrutiny of the second thermodynamic law. In most of the 
present treatments of the cosmological thermodynamic equations of state,  
the entropy in a large expanding sub-volume of the universe remains 
constant. In other words, in a ``big bang'' explosion so strong that 
it staggers the human imagination, not even one small bit of 
entropy was created. This view is in stark contrast to those very tiny 
explosions that have existed in human wars which tend to leave an entropy 
of disorder scattered all over the place.

\begin{figure}[htbp]
\begin{center}
\mbox{\epsfig{file=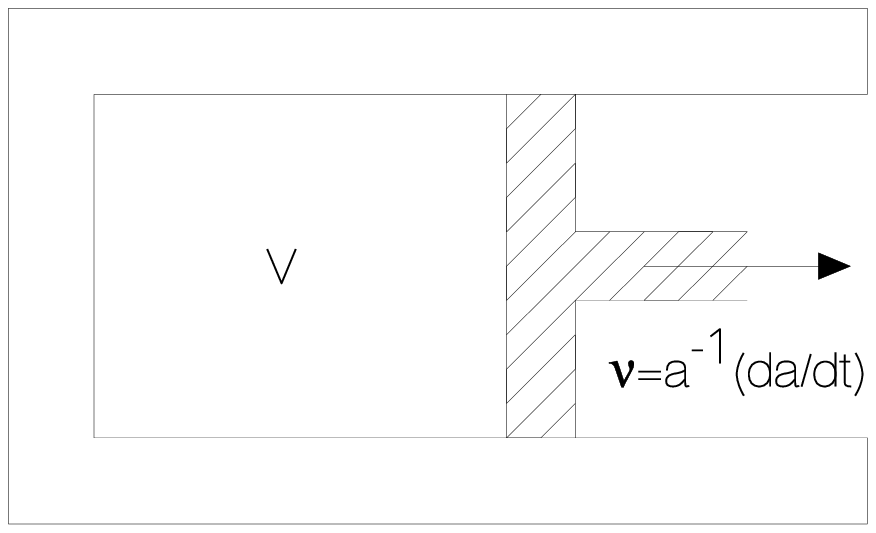,height=70mm}}
\caption{The expansion of the universe at a 
rate $\nu $ is schematically pictured as an expanding sub-volume 
$V$ of the universe here pictured inside a insulating piston. 
The total pressure can be driven negative if the viscous heating of the 
expanding material within the piston is sufficiently large}
\label{grfig1}
\end{center}
\end{figure}

Connected with what we suggest is a failure of the notion of an entropy 
conserving explosion, is the failure of the notion of a completely local 
effective Lagrangian. In the usual standard cosmological model, one may 
derive Eqs.(3) and (4) from an effective local Lagrangian,
which exists whenever the coordinate of interest \begin{math} a(t) \end{math} 
is changing slowly compared with all of the other coordinates which 
are being integrated out of the cosmological dynamics. Since the 
Hubble expansion rate\cite{22,23} 
\begin{equation}
\nu =\left({\dot{a}\over a}\right)
\end{equation}
diverges at the big bang, (\begin{math}\nu\to \infty\end{math} as 
\begin{math}t\to 0\end{math}), the regime in which 
\begin{math} a(t) \end{math} varies {\em slowly} is not abundantly clear.
The mechanism for inflation here being discussed is shown above in FIG.1.
One follows a material sub-volume \begin{math} V \end{math} of the universe, 
here pictured in a schematic fashion as material inside a piston. 
Neither heat nor matter flows into or out of the sub-volume 
\begin{math} V \end{math}.

In Sec.II the notion of an effective Lagrangian for 
\begin{math} a(t) \end{math} will be derived from the action principle 
of general relativity. The gravitational part of the effective Lagrangian 
may be derived from the scalar curvature in the usual manner. The matter 
part of the effective Lagrangian may be computed from the energy per unit 
volume \begin{math} \varepsilon  \end{math} even for the case wherein 
the effective action is non-local in time and depends on the rate entropy 
production. In Sec.III, the thermodynamics laws of cosmology will be 
reviewed. In Sec.IV, the manner in which the
second law may induce inflationary epochs into the cosmological 
evolution will be discussed. In the concluding Sec.V, our mechanism 
for obtaining inflation without recourse to the cosmological term 
will be reviewed. 

\section{Effective Cosmological Action}

The effective action for general relativity with a general metric 
\begin{equation}
-c^2d\tau ^2=g_{\mu \nu}dx^\mu dx^\nu 
\end{equation}
reads 
\begin{equation}
W=W_{gravity}+W_{matter}
\end{equation}
where 
\begin{equation}
W_{gravity}=\left({c^3\over 16\pi G}\right)\int R
\left(\sqrt{-g}d^4x\right)
\end{equation}
is the gravitational action; i.e. \begin{math}R\end{math} is the scalar 
curvature. For the standard cosmological metric in Eq.(1), 
\begin{equation}
R=\left({6\over a^2 c^2}\right)
\left(a\ddot{a}+\dot{a}^2+c^2\kappa \right).
\end{equation} 
Furthermore 
\begin{equation}
\sqrt{-g}d^4x=ca^3d\tilde{V}_\kappa dt ,
\end{equation}
where 
\begin{equation}
d\tilde{V}_\kappa=\sigma_\kappa \sin\theta d\chi d\theta d\phi ,
\end{equation}
is a time independent spatial volume element. The time dependent 
expansion of all such cosmological spatial volume elements is described 
by the scale factor \begin{math}a^3 \end{math} on the right hand side of 
Eq.(11). From Eqs.(9)-(12), the gravitation action of a finite but large 
sub-volume of the universe is given by
\begin{equation}
\tilde{W}_{gravity}=\int L^\prime_{gravity}dt 
\end{equation}
where 
\begin{equation}
L^\prime_{gravity}= 
\left({3c^2\over 8\pi G}\right)
\int_{\tilde{V}_\kappa} \left(a^2\ddot{a}+a\dot{a}^2+c^2a\kappa \right) 
d\tilde{V}_\kappa .
\end{equation} 
The Lagrangian \begin{math}L^\prime_{gravity} \end{math} depends 
on the second derivative \begin{math}\ddot{a} \end{math}. However, 
an integration by parts in Eq.(13) shows that the  
gravitational part of the Lagrangian containing at most the first 
derivative \begin{math}\dot{a} \end{math} may be written 
\begin{equation}
L_{gravity}= 
\left({3c^2\over 8\pi G}\right)
\int_{\tilde{V}_\kappa} \left(c^2a\kappa -a\dot{a}^2\right) 
d\tilde{V}_\kappa , 
\end{equation}
or equivalently
\begin{equation}
\tilde{W}_{gravity}=\int L_{gravity}dt 
\end{equation}
where 
\begin{equation}
L_{gravity}= 
\left({3c^2\tilde{V}_\kappa \over 8\pi G}\right)
\left(c^2a\kappa -a\dot{a}^2\right). 
\end{equation}
Eq.(17) contains a complete description of the purely 
gravitational part of the Lagrangian under the standard 
cosmological hypothesis of a locally isotropic universe. 
From the gravitational part of the Lagrangian one may obtain 
the gravitational part of the energy 
\begin{equation}
{\cal E}_{gravity}=\dot{a}\left({\partial L_{gravity}\over 
\partial \dot{a}}\right)-L_{gravity},
\end{equation}
which reads 
\begin{equation}
{\cal E}_{gravity}=-\left({3c^2\tilde{V}_\kappa \over 8\pi G}\right)
\left(c^2a\kappa +a\dot{a}^2\right). 
\end{equation}
The matter contribution to the Lagrangian is a more 
subtle computation.

To begin, one may consider the pressure-energy tensor of the 
matter \begin{math} T_{\mu \nu}  \end{math}. This second 
rank tensor has one time-like eigenvector 
\begin{math} v^\mu   \end{math}; i.e. 
\begin{equation}
T_{\mu \nu}v^\nu =-\varepsilon v_\mu , 
\ \ \ v_\mu v^\mu =-c^2.
\end{equation}
The eigenvalue \begin{math}\varepsilon  \end{math} represents 
the energy per unit volume while 
\begin{math} v^\mu  \end{math} represents the local 
``fluid'' mechanical velocity field of the matter. The metric may 
now be written in the form 
\begin{equation}
g_{\mu \nu}=-(v_\mu v_\nu /c^2)+h_{\mu \nu}, 
\ \ \ h_{\mu \nu}v^\nu =0,
\end{equation}
so that an observer moving with the matter at velocity 
\begin{math} v^\mu  \end{math} observes a local time interval 
\begin{equation}
c^2 dt^2=(v_\mu dx^\mu /c)^2
\end{equation}  
and a local space interval 
\begin{equation}
dl^2=h_{\mu \nu}dx^\mu dx^\nu 
\end{equation}  
or (equivalently) a proper time 
\begin{equation}
c^2 d\tau ^2=c^2 dt^2-dl^2.
\end{equation}
Eqs.(1) and (24) are made consistent with  
\begin{equation}
dl^2=a(t)^2\left(d\chi^2 +
\sigma_\kappa (\chi )(d\theta ^2+\sin^2\theta d\phi^2)\right).
\end{equation}

Under the standard cosmological hypothesis that the universe 
is isotropic with pressure \begin{math} P  \end{math}, 
the pressure-energy tensor must have the form 
\begin{equation}
T_{\mu \nu}=\varepsilon (v_\mu v_\nu /c^2)+Ph_{\mu \nu},
\end{equation}
more usually written 
\begin{equation}
T_{\mu \nu}=(\varepsilon +P)(v_\mu v_\nu /c^2)+Pg_{\mu \nu}.
\end{equation}
The pressure-energy tensor is related to the matter action via the 
variational equation 
\begin{equation}
\delta W_{matter}=\left({1\over 2c}\right)
\int T^{\mu \nu}\delta g_{\mu \nu }\sqrt{-g}d^4 x. 
\end{equation}
During the course of cosmological expansion 
\begin{equation}
\delta g_{\mu \nu }=\left({2\delta a\over a}\right)h_{\mu \nu},
\end{equation}
so that 
\begin{equation}
\delta W_{matter}=\left({1\over c}\right)
\int P h^{\mu \nu}h_{\mu \nu}\left({\delta a\over a}\right)
\sqrt{-g}d^4 x. 
\end{equation}
Employing Eqs.(11) and (30) as well as  
\begin{math} h^{\mu \nu}h_{\mu \nu}=3  \end{math}, yields the variational 
equation for a finite (but large) sub-volume 
\begin{math} \tilde{W}_{matter}=\int L_{matter}dt  \end{math}; i.e.  
\begin{equation}
\delta L_{matter}=
\int_{\tilde{V}_\kappa }\left({3\delta a \over a}\right)
Pa^3d\tilde{V}_\kappa ,
\end{equation}
which integrates to 
\begin{equation}
\delta L_{matter}=
\left({3\delta a \over a}\right)
Pa^3\tilde{V}_\kappa .
\end{equation}

At this point one may employ the pressure-energy tensor identity 
\begin{math} D_{\mu }T^{\mu \nu }=0  \end{math}, or more simply 
compute the work done on the sub-volume during the expansion 
\begin{equation}
\delta {\cal E}_{matter}=-P\delta V=\delta (V\varepsilon ),
\ \ \ (V=a^3\tilde{V}_\kappa ).
\end{equation}
Hence 
\begin{equation}
\delta \varepsilon =-(\varepsilon +P)\left({\delta V\over V}\right)
=-(\varepsilon +P)\left({3\delta a\over a}\right).
\end{equation}
From Eqs.(32) and (34)
\begin{equation}
\delta L_{matter}=-
\left\{\delta \varepsilon +
\left({3\delta a\over a}\right)\varepsilon \right\}
a^3 \tilde{V}_{\kappa }.
\end{equation}
From Eq.(35) one proves the following \medskip \\  
{\bf Theorem 1:} {\em The 
matter Lagrangian for a large sub-volume 
{\it V} of the universe 
is the negative of the matter energy contained within that sub-volume, }
\begin{equation}
L_{matter}=-\varepsilon V=-\varepsilon a^3\tilde{V}_\kappa 
=-{\cal E}_{matter}.
\end{equation}
The theorem holds true for any model with a local or non-local 
matter Lagrangian and/or when internal heating produces entropy.
Finally if one adds the gravitational energy to the matter energy 
the resulting total energy is zero; i.e.  
\begin{equation}
{\cal E}={\cal E}_{gravity}+{\cal E}_{matter}=0.
\end{equation}
From Eqs.(19) and (36), one obtains the conventional cosmological 
equation
\begin{equation}
\dot{a}^2+c^2\kappa =\left({8\pi G\over 3c^2}\right)a^2\varepsilon .
\end{equation}

Note that the classical general relativity condition of zero total 
energy, for every sub-volume in Eq.(37), is often written in quantum 
mechanical terms employing the Schr\"odinger equation 
\begin{math}{\cal H}\Psi ={\cal E}\Psi \end{math} for the wave function 
\begin{math} \Psi \end{math}. Since 
\begin{math} {\cal E}=0 \end{math}  the ``wave function of the 
universe'' \begin{math} \Psi \end{math} obeys  
\begin{math}{\cal H}\Psi =0 \end{math}. 
We shall not compute \begin{math} \Psi \end{math} in detail.

\section{Cosmological Thermodynamics}

In the standard cosmological model it assumed that the total entropy 
\begin{math} {\cal S} \end{math} in a large sub-volume 
\begin{math} V=a^3\tilde{V}_\kappa  \end{math} is zero. For a 
quasi-static yet irreversible cosmological expansion, the entropy 
\begin{math} S \end{math} increases with time. In thermodynamic 
terms, the energy of a large sub-volume obeys 
\begin{equation}
d{\cal E}_{matter}=Td{\cal S}-pdV,
\end{equation} 
where \begin{math} p \end{math} is the {\em thermodynamic} pressure. 
If the expanding sub-volume yields internal heating from {\em viscous}  
pressure, then the thermodynamic pressure 
\begin{math} p \end{math} cannot be identified with the 
total pressure \begin{math} P \end{math} in the pressure-energy tensor. 
With purely internal heating, there is no heat {\em flow} into the 
expanding sub-volume. One may employ the energy Eq.(33) 
\begin{equation}
d{\cal E}_{matter}=-PdV,
\end{equation}
relating the total pressure \begin{math} P \end{math} to the 
thermodynamic pressure \begin{math} p \end{math} via  
the internal heating rate \begin{math} \dot{Q}=T\dot{\cal S} \end{math}.  
From Eqs.(39) and (40) one proves the following \medskip \\  
{\bf Theorem 2:} {\em The ratio of the heating rate 
\begin{math} \dot{Q} \end{math} to the volume expansion 
rate \begin{math} \dot{V} \end{math} subtracts from the 
thermodynamic pressure to yield the total pressure, }  
\begin{equation}
P=p-\left({\dot{Q}\over \dot{V}}\right)
\end{equation}
While the thermodynamic pressure \begin{math} p \end{math} may appear to  
be positive, the total pressure \begin{math} P \end{math} may go negative 
if there is sufficient internal heating. If 
\begin{math} \dot{Q}>p\dot{V}  \end{math}, 
then  \begin{math} P<0  \end{math}. Note that one need {\em not} 
introduce an inflaton field to achieve a negative total pressure. 
Simple internal heating can do the job.  

Eqs.(39) and (40), with an entropy per unit volume of 
\begin{math} s=({\cal S}/V) \end{math}, yield
\begin{equation}
d(V\varepsilon )=-PdV=Td(Vs)-pdV,
\end{equation} 
which implies 
\begin{equation}
d\varepsilon =-(\varepsilon +P)\left({dV\over V}\right),
\end{equation}
and
\begin{equation}
d\varepsilon = -(\varepsilon +p-Ts)\left({dV\over V}\right)+Tds.
\end{equation} 
 
\section{Entropy Production} 

As stated in Sec.I, it is unreasonable to assume for the largest 
explosion ever conceived (i.e. the big bang) that the internal 
heating (i.e. entropy production) should be strictly zero. During 
a universal expansion one may define a ``longitudinal viscosity'' 
\begin{math} \zeta  \end{math}. In detail, the fact that near by 
matter moves away from us at a velocity 
\begin{math}{\bf v}=(\dot{a}/a) {\bf r} \end{math} yields a viscous 
pressure  
\begin{equation}
p_{viscous}=-\zeta div{\bf v}=-3\zeta \left({\dot{a}\over a}\right)
=-\left({\dot{Q}\over \dot{V}}\right).
\end{equation}
in accordance with Eq.(41). Using the identity 
\begin{math}(\dot{V}/V)=3(\dot{a}/a)  \end{math}
yields the heating rate per unit volume of the universal expansion, 
\begin{equation}
\left({\dot{Q}\over V}\right)=9\zeta \left({\dot{a}\over a}\right)^2.
\end{equation}
Furthermore, from Eqs.(4), (41) and (45) one finds 
\begin{equation}
\Omega^2=\left({4\pi G\over 3 c^2}\right)
\left\{\varepsilon +3p-9\zeta \left({\dot{a}\over a}\right)\right\}.
\end{equation}

The usual parameters employed in the standard cosmology are 
(i) the Hubble expansion rate \begin{math} \nu \end{math} and 
(ii) the universal ``deacceleration'' \begin{math} \phi  \end{math}; 
i.e.
\begin{equation}
\nu =\left({\dot{a}\over a}\right),\ \ \ 
\phi =-\left({a\ddot{a}\over \dot{a}^2}\right).
\end{equation} 
Alternatively \begin{math}\phi = (\Omega /\nu )^2 \end{math}, 
leading to the following \medskip \\ 
{\bf Theorem 3:} {\em The deacceleration parameter 
\begin{math} \phi  \end{math} in a cosmology which 
includes entropy production may be written}
\begin{equation}
\phi =\left({4\pi G\over 3 \nu^2 c^2}\right)
\left(\varepsilon +3p-9\nu \zeta \right).
\end{equation}
Eq.(49) is the central result of this work. An inflationary epoch 
of a standard cosmology will exist if the longitudinal viscosity 
is sufficiently large 
\begin{equation}
9\nu \zeta >
\left(\varepsilon +3p\right)\ \ 
{\rm implies}\ \ \phi <0,
\end{equation} 
where \begin{math} p \end{math} is the thermodynamic 
pressure. The condition \begin{math} \phi <0  \end{math} 
also {\em defines inflation} if \begin{math} |\phi |>>1  \end{math} 
and {\em defines quintessence} if  
\begin{math} |\phi |<<1  \end{math}. The theory of inflation 
and quintessence can then be based on the longitudinal viscosity 
without recourse to the introduction of new fields invented 
merely for cosmological considerations.

In order to estimate the viscous mechanism for inflation, one may start 
from the Kubo formula for longitudinal viscosity in terms of pressure 
fluctuations \begin{math} \Delta P \end{math} at temperature 
\begin{math} T \end{math}; 
$$
\zeta =\left(1\over k_B TV\right)\times 
$$
\begin{equation} 
\int_0^\infty dt 
\int_V d^3{\bf r}  \int_V d^3{\bf r}^\prime 
{\Re }e\left<\Delta P({\bf r},t)\Delta P({\bf r}^\prime ,0)\right>.
\end{equation} 
On the other hand, the thermodynamic adiabatic compressibility, 
\begin{equation}
K_s=-\left({1\over V}\right)
\left({\partial V\over \partial p}\right)_{\cal S},
\end{equation}
is determined by the {\em static} pressure fluctuations at a 
given time;  
\begin{equation}
K_s^{-1}=\left(1\over k_B TV\right)
\int_V d^3{\bf r}  \int_V d^3{\bf r}^\prime 
\left<\Delta P({\bf r})\Delta P({\bf r}^\prime )\right>.
\end{equation}
Hence, if \begin{math} \tau_p  \end{math} denotes the relaxation 
time for pressure fluctuations, then one obtains 
\begin{equation}
\zeta =\tau_p K_s^{-1}.
\end{equation}
From Eqs.(49) and (54), the final expression for the deacceleration 
parameter reads 
\begin{equation}
\phi =\left({4\pi G\over 3 \nu^2 c^2}\right)
\left\{\varepsilon +3p-\left({9\nu \tau_p \over K_s}\right)\right\}.
\end{equation}
Eq.(55) is in a form suitable for estimating the conditions for which 
inflation \begin{math} \phi<0  \end{math} holds true.

\section{Conclusions}

The sign of the deacceleration parameter 
\begin{math} \phi \end{math} is completely determined by Eq.(55). 
An inflationary period is described by the condition  
\begin{math} \phi <0 \end{math}. In the``big bang'' limit 
\begin{math} t\to 0 \end{math}, one expects very high particle energies 
on the scale of the particle rest energy; i.e. 
\begin{math} \epsilon({\bf p})>>mc^2  \end{math} which is certainly 
true for massless particles such as photons or neutrinos. In such a 
big bang high energy limit, the energy density 
\begin{math} \varepsilon \end{math} is related 
to the thermodynamic pressure \begin{math} p \end{math} via 
\begin{math} \varepsilon \approx 3p \end{math}. Thermodynamics then 
yields an inverse adiabatic compressibility of 
\begin{math} K_s^{-1}\approx (4\varepsilon/9) \end{math}. Thus, 
near the big bang 
\begin{equation}
\phi \to \left({8\pi G\varepsilon \over 3 \nu^2 c^2}\right)
(1-2\nu \tau_p)\ \ {\rm as}\ \ t\to 0.
\end{equation}
The condition for inflation {\em near the big bang} then reads 
\begin{equation}
\tau_p >\left({1\over 2\nu }\right) \ \ {\rm as}\ \ t\to 0
\ \ {\rm (inflation)}.
\end{equation}

To see what is involved, the inverse Hubble expansion rate 
\begin{math} \nu^{-1}  \end{math} is roughly the age of the 
new born universe. It is evident that when a violent explosion 
first begins, the time scale \begin{math} \tau_p  \end{math} 
for the pressure to reach equilibrium is long compared to  
a very short time scale for a young explosion. If the  {\em pressure 
disequilibrium of the big bang} lasts longer than the age of a  
new-born universe(\begin{math} \tau_p>>\nu^{-1}\end{math}), 
then an early time period for inflation would exist. 
Inflaton fields need not play any role in the process. 

A present day inflation, albeit quite small, is somewhat more difficult 
to explain. In a matter dominated universe 
\begin{math} \varepsilon\approx \rho c^2>>p  \end{math}, 
while the adiabatic compressibility 
\begin{math} K_s\approx (\rho u_s^2)^{-1}  \end{math} where 
\begin{math} \rho  \end{math} is the mass density and 
\begin{math} u_s  \end{math} is the adiabatic sound velocity. 
In the matter dominated present, Eq.(55) reads
\begin{equation}
\phi \approx \left({4\pi \rho G\over 3 \nu^2 }\right)
\left\{1-\left({u_s^2\nu \tau_p\over c^2}\right)\right\}.
\end{equation} 
The present universe is thereby considered to be a ``dilute gas'' 
in a sense to be discussed below.

Let us first consider a dilute gas of slowly moving (compared to light 
speed) molecules in a laboratory volume \begin{math} V \end{math} 
controlled by a piston (similar to FIG.1). The mean square center of 
mass velocity \begin{math} \left<{\bf v}^2 \right>=(3k_BT/M) \end{math} 
of a molecule of mass \begin{math} M \end{math} is determined by the 
temperature \begin{math} T \end{math}. And as volume 
\begin{math} V \end{math} of the gas 
is increased, the mean center of mass kinetic energy of the molecule 
gets smaller. However, the mean {\em internal energy} of the molecules 
does not necessarily cool as fast as the translational center of mass 
kinetic energy of the molecules. There must be many collisions 
between molecules to achieve the local thermal equilibrium 
equipartition of translational kinetic energy and vibrational 
or rotational internal energy of the molecules. The delay time for 
this internal energy equilibrium determines the longitudinal viscosity 
\begin{math} \zeta  \end{math} of the gas.

In the present universe there exists a ``dilute gas of galaxies'', 
i.e. rather complicated ``molecules'' with very many degrees of 
freedom. The galaxies move at velocities of (say) 
\begin{math} (u_s/c)\sim 10^{-3}  \end{math}. Hence, from Eq.(58) 
we see that there can be a present day negative deacceleration if (say)  
\begin{math} \nu \tau_p > 10^{-6}  \end{math}. In other words, the 
universal expansion forces the center of mass momentum 
\begin{math} {\bf p} \end{math} of a galaxy to slow down according 
to the cosmological friction law 
\begin{math} d{\bf p}/dt)=-\nu {\bf p} \end{math}. But the internal energies 
of the galaxies are not in equipartition thermal equilibrium with the 
translational kinetic energy of a galaxy. (If such 
equilibrium did exist we would all be dead.) For such equilibrium to 
exists we would need several collisions between galaxies converting 
translational energies to internal energies. If less that one in a million 
galaxies have undergone collisions, then it is feasible that the resulting 
viscous heating would be sufficient to cause 
\begin{math} \phi <0  \end{math}. The argument for a small present day 
inflation is (of course) not as theoretically strong as the 
argument for inflation at times shortly after the big bang.


\begin{thebibliography}{99}

\bibitem{1}R.C. Tolman, {\it Relativity Thermodynamics and
Cosmology}, Dover Publications Inc., New York (1987).

\bibitem{2} L.D.landau and E.M.Lifshitz, {\it The Classical Theory of
Fields}, Pergamon press Ltd., New York (1975).

\bibitem{3} C.W.Misner, K.S.Thorne and J.A.Wheeler, {\it Gravitation}, 
W.H.Freeman and Company, New York (1970).

\bibitem{4} S.Weinberg, {\it Gravitation and Cosmology}, John Wiley and
Sons Inc., New York (1972).

\bibitem{5} R.R.Caldwell, R.Dave and P.J.Steinhardt, 
{\it Phys. Rev. Lett.} {\bf 80}, 1582 (1988).

\bibitem{6} P.J.E.Peebles and A.Vilenkin, 
{\it Phys. Rev.} {\bf D59}, 063505-1 (1999).

\bibitem{7} S.W.Hawking, {\it Phys. Rev.} {\bf D37}, 904 (1988).

\bibitem{8} S. Coleman, {\it Nucl. Phys.} {\bf B310}, 643 (1988).

\bibitem{9} P.J.E. Peebles and B.Ratra, {\it Astro. Phys.J.352}, 
L171 (1988).

\bibitem{10} S. Weinberg, {\it Rev. Mod. Phys.} {\bf 61}, 1 (1989).  

\bibitem{11} A.H.Guth, {\it Phys. Rev.} {\bf D23}, 347 (1981).

\bibitem{12} A.Vilenkin, {\it Phys. Rev.} {\bf D27}, 2848 (1983).

\bibitem{13} L.H.Ford, {\it Phys. Rev.} {\bf D35}, 2955 (1987).

\bibitem{14} A.Linde, {\it Phys. Lett.} {\bf B238}, 160 (1990).

\bibitem{15} G. Veneziano, {\it Phys. Lett.} {\bf B265}, 287 (1991).

\bibitem{16} B.Spokoiny, {\it Phys. Lett.} {\bf B315}, 40 (1993).

\bibitem{17} M.Gasperini and G.Veniziano, {\it Astropart.Phys.J}, 317 (1993).

\bibitem{18} G. Veneziano, {\it Phys. Lett.} {\bf B406}, 297 (1997).

\bibitem{19} J.E. Lidsey, A.R.Liddle, E.W. Kolb, E.J. Copeland, 
T. Harreiro and M. Abney, {\it Rev. Mod. Phys.} {\bf 69}, 373 (1997).

\bibitem{20} K.I. Izawa, M. Kawasaki, and T. Yanagida, 
{\it Phys. Lett.} {\bf B411}, 249 (1997).

\bibitem{21} D.M. Lyth and E.D. Stewart {\it Phys. Rev.} 
{\bf D53}, 1784 (1996).

\bibitem{22} A.G. Riess, W.H.Prees, and R.P.Kirshner, 
{\it Astrophys. J. Lett.438}, 17(1995)

\bibitem{23} T. Kundic, et. al. {\it Astrophys.J.482}, 75 (1997).

\end{thebibliography}
\end{document}